\newif\ifpreprint
\newif\ifdraft

\draftfalse
\preprinttrue

\documentclass[twocolumn, switch]{article}
\usepackage{preprint}

\usepackage{xcolor}
\usepackage{url}            
\usepackage{datetime}       
\usepackage{booktabs}       
\usepackage{amsfonts}       
\usepackage{nicefrac}       
\usepackage{microtype}      
\usepackage{graphicx}
\usepackage{doi}
\usepackage[labelfont=bf,small]{caption}
\usepackage{etoolbox}
\usepackage{geometry}
\usepackage{tabularx} 
\usepackage{float}
\usepackage{placeins}
\usepackage{glossaries}

\usepackage{booktabs}
\usepackage{graphicx}
\usepackage{url}

\usepackage[utf8]{inputenc}	
\usepackage[T1]{fontenc}
\usepackage{xcolor}
\usepackage{lineno}					
\usepackage{tikz} 					

   \usepackage{hyperref}
    \hypersetup{colorlinks=true, linkcolor=teal, urlcolor=teal, citecolor=teal, anchorcolor=teal}
    \usepackage[numbers,square]{natbib}
\usepackage{titlesec}
\titlespacing\section{0pt}{12pt plus 3pt minus 3pt}{1pt plus 1pt minus 1pt}
\titlespacing\subsection{0pt}{10pt plus 3pt minus 3pt}{1pt plus 1pt minus 1pt}
\titlespacing\subsubsection{0pt}{8pt plus 3pt minus 3pt}{1pt plus 1pt minus 1pt}

\definecolor{lime}{HTML}{A6CE39}
\DeclareRobustCommand{\orcidicon}{
	\begin{tikzpicture}
	\draw[lime, fill=lime] (0,0)
	circle [radius=0.16]
	node[white] {{\fontfamily{qag}\selectfont \tiny ID}};
	\draw[white, fill=white] (-0.0625,0.095)
	circle [radius=0.007];
	\end{tikzpicture}
	\hspace{-2mm}
}

\ifdraft
    \usepackage{xwatermark}
    \newwatermark[allpages,color=gray!60,angle=90,scale=0.32, xpos=-4.05in,ypos=0]{DRAFT MANUSCRIPT - DO NOT SHARE}
    \newwatermark[allpages,color=gray!60,angle=90,scale=0.32, xpos=3.9in,ypos=0]{DRAFT MANUSCRIPT - HAS NOT BEEN PEER-REVIEWED}
    \newwatermark[allpages,color=gray!90,angle=0,scale=0.28, xpos=0in,ypos=-5in]{DRAFT MANUSCRIPT}

    \usepackage{draftwatermark}
    \SetWatermarkText{DRAFT}
    \SetWatermarkScale{0.8}
    \SetWatermarkColor[gray]{0.95}
\fi

\title{Synthetic Patients: Simulating Difficult Conversations with Multimodal Generative AI for Medical Education}
\shorttitle{Synthetic Patients}

\usepackage{authblk}

\author[1]{Simon N Chu, MD, MS\href{https://orcid.org/0000-0003-3863-1548}{\orcidicon}}
\author[2]{Alex J Goodell, MD, MS\href{https://orcid.org/0000-0003-0229-8843}{\orcidicon}}

\affil[1]{Department of Surgery, School of Medicine, University of California, San Francisco}
\affil[2]{Anesthesia Informatics and Media Lab, Department of Anesthesiology, Pain, and Perioperative Medicine, School of Medicine, Stanford University}

\begin{document}

\twocolumn[\begin{@twocolumnfalse}
\maketitle

\begin{abstract}
\textbf{Problem:}
Effective patient-centered communication is a core competency for physicians. However, both seasoned providers and medical trainees report decreased confidence in leading conversations on sensitive topics such as goals of care or end-of-life discussions. The significant administrative burden and the resources required to provide dedicated training in leading difficult conversations has been a long-standing problem in medical education.
\textbf{Approach:}
In this work, we present a novel educational tool designed to facilitate interactive, real-time simulations of difficult conversations in a video-based format through the use of multimodal generative artificial intelligence (AI). Leveraging recent advances in language modeling, computer vision, and generative audio, this tool creates realistic, interactive scenarios with avatars,  or “synthetic patients.” These synthetic patients interact with users throughout various stages of medical care using a custom-built video chat application, offering learners the chance to practice conversations with patients from diverse belief systems, personalities, and ethnic backgrounds. 
\textbf{Outcomes:}
While the development of this platform demanded substantial upfront investment in labor, it offers a highly-realistic simulation experience with minimal financial investment. For medical trainees, this educational tool can be implemented within programs to simulate patient-provider conversations and can be incorporated into existing palliative care curriculum to provide a scalable, high-fidelity simulation environment for mastering difficult conversations. 
\textbf{Next Steps:}
Future developments will explore enhancing the authenticity of these encounters by working with patients to incorporate their histories and personalities, as well as employing the use of AI-generated evaluations to offer immediate, constructive feedback to learners post-simulation.
\end{abstract}
\vspace{0.5cm}

\end{@twocolumnfalse}]


\begin{figure*}[!htb]
\centering
\includegraphics[width=0.7\linewidth]{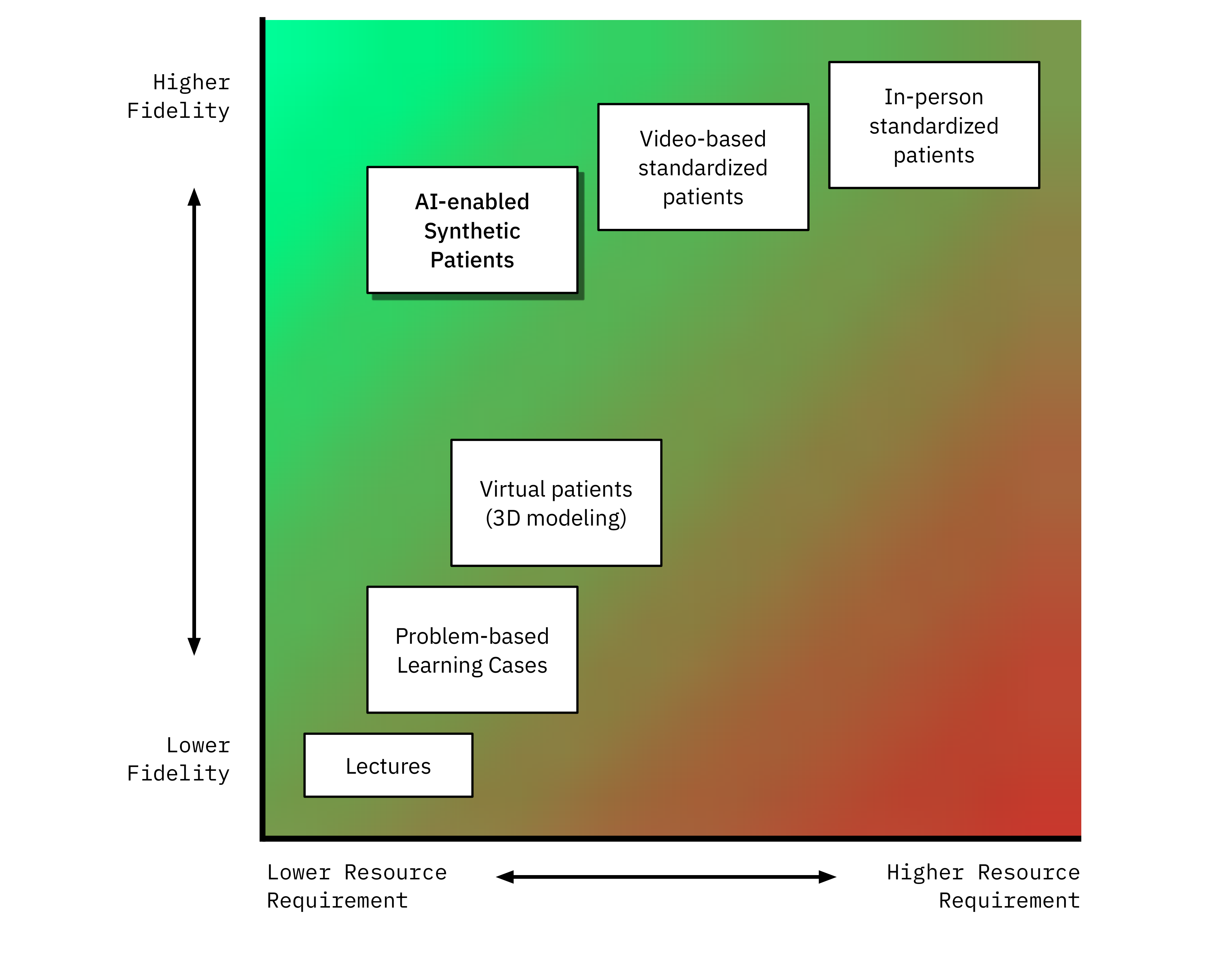}
\begin{minipage}[b]{0.8\textwidth}
    \vspace{3mm} 
     \caption{\small \textbf{Overview of medical training modalities for teaching difficult conversations.} 
     Schematic illustrating current medical training modalities utilized for teaching difficult conversation skills, positioned according to their fidelity and resource requirements. High fidelity but high resource requirement modalities include in-person standardized patients and video-based standardized patients. Low fidelity and low resource demanding methods include lectures and problem-based learning cases. AI-enabled synthetic patients deliver a high-fidelity experience with relatively lower resource requirements, potentially offering an optimal balance for effective training with minimal implementation resources.}
    \label{2d_matrix}
   \end{minipage}
\end{figure*}

\section{Problem}

Healthcare providers routinely engage in difficult conversations with patients and their families, discussing topics such as unexpected diagnoses, personal sexual behaviors, or sensitive mental health issues. These conversations, when performed well, can have a significant impact on patient experience, adherence to treatment, and overall health outcomes. One particularly challenging area concerns discussing a patient’s goals of care, including their wishes for or against resuscitation efforts. Unfortunately, both seasoned providers and trainees alike report a lack of confidence in their ability to facilitate these discussions, and trainees across various specialties express a consistent need for additional training in this domain \cite{Schmit2016}. Studies have shown that only about half of trainees felt comfortable leading goals of care conversations \cite{Bonanno2019}. Despite both trainees and faculty acknowledging the need for additional training in this area, significant barriers, including constrained trainee time and a shortage of educational resources, hinder the widespread integration of communication skills curricula into medical school or residency curriculum \cite{Bonanno2019}. Currently, the predominant teaching method involves experiential “on-the-job” training supplemented by dedicated lectures \cite{Allar2022}. This conventional approach, characterized by its passive educational strategies, may not provide the necessary interactive and dedicated training for effective practice in sensitive conversations. Consequently, it may fall short in adequately preparing trainees, thereby undermining their confidence in effectively managing difficult conversations.

Several strategies have been proposed to address this problem, including the development of tailored educational resources, interactive modules, or simulation-based experiences using standardized patient actors\cite{Benesch2022,Yuen2013}. Although simulation-based experiences have been shown to boost residents’ self-perceived confidence, these scenarios are resource intensive and difficult to scale. They necessitate the use of paid actors, require scheduling protected time for both trainees and facilitators, and demand substantial administrative funding and programmatic support \cite{Nagpal2021}. Subsequently, a variety of lower-fidelity, cost-sensitive approaches have been explored, including simulated video visits with patient actors (Figure \ref{2d_matrix}). More recently, the use of virtual patients has been explored. In these simulations, a rendering of a patient is made using 3D modeling technologies. While these approaches reduce the logistic overhead of simulation, their effectiveness has been compromised by low fidelity. Students often report that these models lack emotional and interpersonal depth, and struggle with a lack of authenticity to the interaction with “badly done” or “static” avatars \cite{Kang_2023}.

Recent advances in artificial intelligence may offer a superior approach compared to either traditional actor-based simulations or virtual patients. Since late 2022, large language models, such as GPT4, have been shown to convincingly mimic human text-based conversations. These improvements have prompted some to explore the use of chatbots as simulation tools for early medical trainees, as well as propose using focused models to simulate or provide feedback on difficult conversations \cite{Scherr2023, Burry2024}. Concurrently, rapid developments in computer vision and generative audio have made it possible to create unique images, videos, and voices with publicly-available tools. Collectively, the culmination of these advances in multiple domains (i.e., “multimodal AI”)  has led to development of conversational avatars with realistic speech and facial expressions. Though these avatars have been deployed in commercial domains, to our knowledge, they have not yet been implemented in the realm of medical simulation. 

\begin{figure*}[!htb]
\centering
\includegraphics[width=1\linewidth]{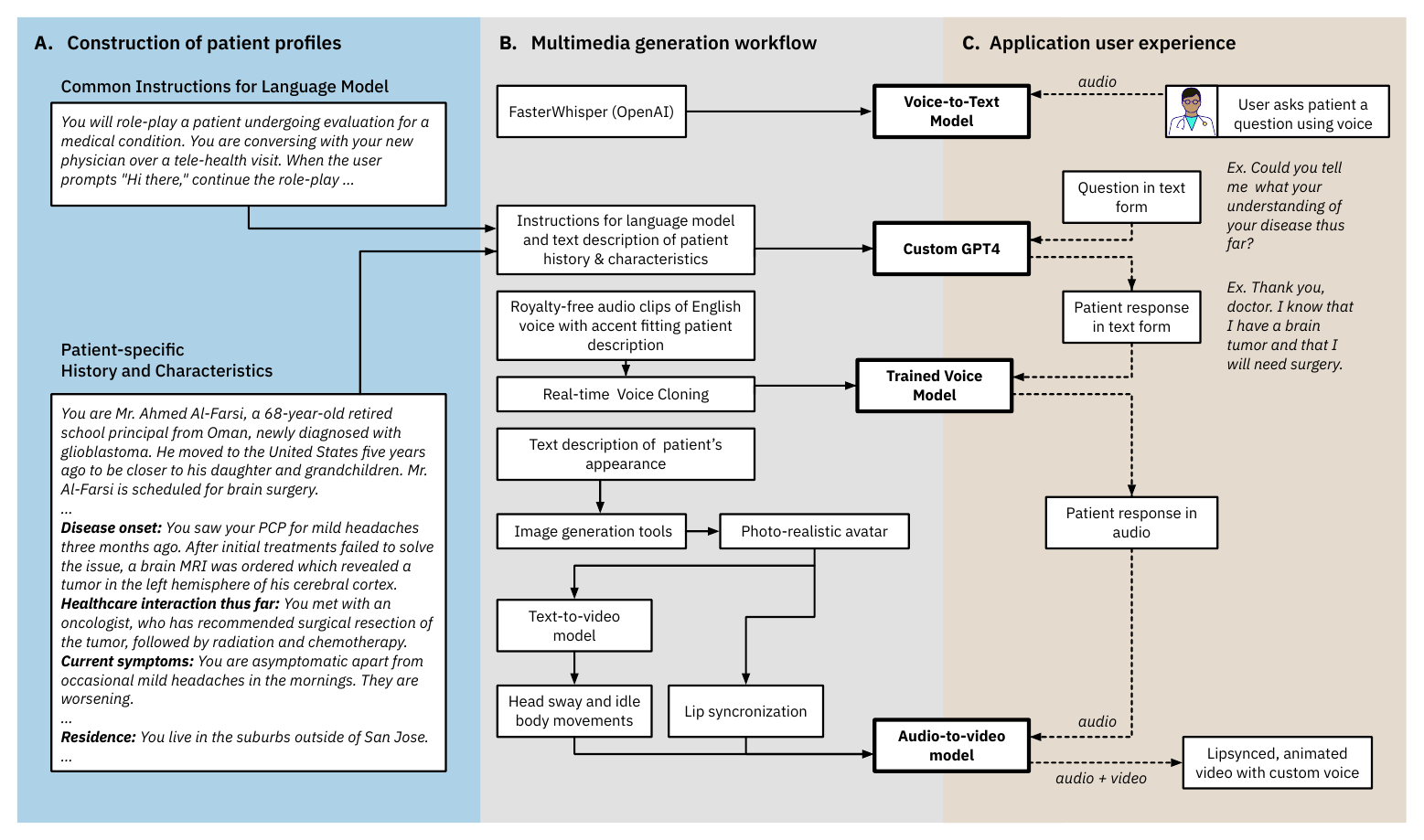}
\begin{minipage}[b]{1\textwidth}
    \vspace{3mm} 
     \caption{\small \textbf{Approach — Overview of synthetic patient generation workflow and user experience on web-based platform.}
     \textbf{A}. To construct of patient profiles, a standardized set of role-playing instructions was given to the language model. Each chatbot received individual patient profiles detailing characteristics like disease onset, healthcare experience, and belief system. Responses were then validated to ensure they aligned with the synthetic patient’s defined attributes and tone. \textbf{B}. Schematic showing how the various tools, including image editors, voice cloning software, and text-to-video generators, were used to create image, audio, and video multimedia, crafting a realistic patient avatar that aligned with the synthetic patient profile. Lines represent the flow of data as input and/or outputs to various tools. \textbf{C}. Schematic showing the flow of the user experience in the web-based platform (dotted lines) and data (solid lines) to generate the media sent to the user. This system allows users to voice questions and receive synchronized audiovisual responses.}
    \label{technical_flow_diagram}
   \end{minipage}
\end{figure*}

\section{Approach}

In this study, we present a novel approach to simulating difficult conversations using AI-generated avatars complete with a unique appearance, personality, and voice. Unlike prior generations of virtual patients, these avatars offer an unprecedented realism and richness of conversation. To differentiate these approaches, we have referred to these AI avatars as “synthetic patients,” to align with the terminology used in other recent machine learning advances. 

This intervention involved three primary steps: (1) creation of patient profiles, (2) generation of patient multimedia, and (3) integration with a custom video chat application (\autoref{technical_flow_diagram}).

\subsection{Construction of patient profiles}

To create the cohort of synthetic patients, we first developed a set of fictional personas depicting individuals at various stages of healthcare interaction, each chosen to illustrate a common encounter involving a difficult conversation. After formulation of a basic text description of the patient and their medical history, we built upon this initial framework by adding individual characteristics, such as a values and belief system, understanding of disease, and personality traits. The patient's profile was then combined with a set of common role-play instructions and provided to GPT4 (OpenAI, San Francisco, CA), thus launching a chatbot roleplaying as the given patient. The provided profile informed the chatbot’s behavior, responses, and mannerisms. Through multiple design-build-test cycles, the text of the profile was iteratively improved, with the goal of generating responses that were concordant with the synthetic patient’s assigned attributes and tone. 

\begin{figure*}[!htbp]
\centering
\includegraphics[width=1\linewidth]{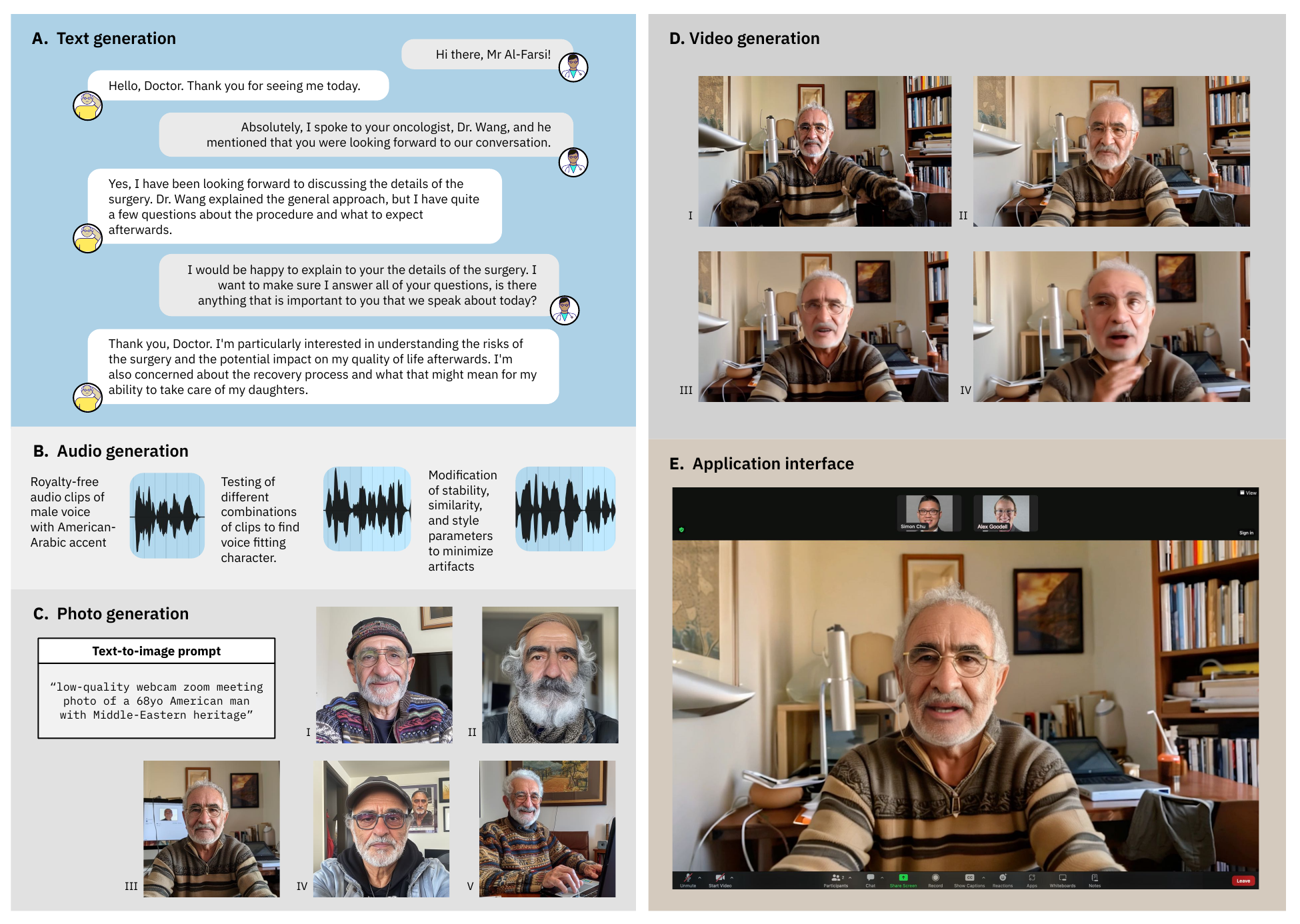}
\begin{minipage}[b]{1\textwidth}
    \vspace{3mm} 
     \caption{\small \textbf{Outcome — Successes and challenges of synthetic patient development.} 
     \textbf{A}. Visual depiction of conversation taken from text-only chatbot. Though occasionally veered off topic, synthetic patients generally offered realistic responses and thoughtful questions. \textbf{B}. Voices for synthetic patients were developed by collecting royalty-free speech clips, refining them with audio processing software, and cloning them using a voice cloning tool. \textbf{C}. Photos of the synthetic patients were generated using multiple imaging models. Encountered problems included overly-stereotypical appearance (images I, II),  duplicated elements (III, IV) and poor understanding of desired perspective (V). \textbf{D}. To generate patient response videos, initial images were processed through video-generation tools to simulate realistic body and head movements. However, these tools introduced hallucinated movements (image I, IV) or significantly distorted the facial features (II, III). \textbf{E}. To enable real-time interaction with synthetic patients, a simple web application was developed which features a mock telehealth interface with a video feed of the patient awaiting a question. The video is comprised of a mix of the above tools and lip-syncing software to provide a real-time realistic telehealth simulation.}
    \label{outcomes_figure}
   \end{minipage}
\end{figure*}
\subsection{Generation of patient multimedia}

To generate realistic images and videos of each synthetic patient, we used multiple generative imaging models (\hyperref[technical_flow_diagram]{Figure 2B}). We aimed to depict the patient at home, engaged in a telehealth video conference. To create a unique voice for the synthetic patient, clips of recorded speech that matched the patient profile were identified, combined, and used as input to a voice-cloning tool to generate a distinctive voice model. Technical implementation details, including specific software used, are available in \hyperref[appendix1]{Appendix 1}.

\subsection{Integration with a custom video chat application}
To facilitate interaction with the synthetic patients, we developed a web-based application that incorporated the patient’s image, voice and text responses into an integrated telehealth simulation, whereby users can ask patients questions using their voice and patients respond with a seamless audiovisual response (Figure \ref{technical_flow_diagram}).

\section{Outcomes}
In this work, we aimed to develop a simulation tool for difficult conversations that combines the authenticity of traditional standardized patients with the flexibility and cost-effectiveness of virtual patients. 

\subsection{Fidelity}
With regard to the fidelity of the simulation, feedback has been widely positive. A video demonstration is available as \hyperref[appendix2]{Appendix 2}. Our approach delivers life-like conversations and realistic videos, effectively simulating telehealth encounters with high fidelity (Figure 3). Initial user experiences with the platform indicate that the interactions are engaging, featuring patients who provide nuanced responses and pose relevant follow-up questions. The spoken audio includes dynamic emotional variations, enhancing empathy with the patient’s situation and bringing the characters to life. 

\subsection{Cost}
Despite the advanced technology underlying our application, development of the tool was inexpensive. Direct costs were approximately \$150. Ongoing costs for running this ensemble of systems would lie primarily in fees associated with hosting the application on a server, approximately \$500-2000 per month depending on the number of users that might be expected to utilize the service. Considering the opportunity cost incurred to coordinate traditional standardized patients, our approach may still be cost-saving. 

Though direct costs were minimal, development of the individual components of the system was time-intensive and required repeated iteration and experimentation by the authors to achieve the desired output. Unlike many prior technologies, use of generative AI requires no specialized programming knowledge, and many of these models are freely available online. For this reason, we expect educators will utilize them extensively in the coming years, and feel it is worthwhile to describe some of the challenges we encountered. 

\subsection{Challenges}
In developing the patient profiles and common instructions for the language model component, while the majority of conversations provided realistic dialogue, some occasionally strayed from the clinical context. These problems were ameliorated by revising patient profiles to be more explicit in instruction, resulting in fluid, realistic dialog (\hyperref[outcomes_figure]{Figure 3A}). Conversely, a variety of challenges were apparent in the generation of patient multimedia. For instance, in creating patients’ voices, the use of cloned real voices combined from multiple sources occasionally resulted in artifactual noises, stuttering, and uneven dictation. Manual experimentation with available parameters mitigated but did not entirely resolve these issues (\hyperref[outcomes_figure]{Figure 3B}). Similarly, the initial images often featured overly-stereotypical representations of patients based on their demographics, raising concern that our examples could reinforce preexisting ethnicity-based tropes (\hyperref[outcomes_figure]{Figure 3C}). Imaging models also struggled to understand the desired perspective of a video conference image and included distracting duplicates of the request – a sort of “visual stuttering” that occasionally required the use of traditional photo-editing software to fix. A multitude of new issues also arose when attempting to bring these images to life with animation. In some cases, the patient would engage in “hallucinated” requests, such as raising their hands in the air, despite no inclusion in the prompt (\hyperref[outcomes_figure]{Figure 3D}). More often, video generation tools distorted the facial structure of the patient during the animation process, resulting in representations that were unrecognizable at best and severely disfigured at worst. A mixture of video components and lip-synchronization was used to develop the user-facing application interface (\hyperref[outcomes_figure]{Figure 3E}).

In summary, though our tool offers high fidelity simulation at a reasonable cost, significant challenges were encountered in developing the resources required for a patient encounter, which may significantly hinder widespread adoption of these systems at this time. 

\section{Next Steps}
While this study presents a technical innovation for a focused use case in medical education, it portends much broader implications. Our platform, or a similar generative AI workflow, can be employed to create a multitude of clinical experiences with innumerable permutations in the patient’s demographic characteristics, values, or disease states. Furthermore, future models could incorporate inputs from real patients and their lived experiences, significantly enhancing the realism of simulated synthetic patient responses. Due to the relatively low cost of developing new scenarios, it may be feasible to integrate this technology into everyday clinical practice. In this way, providers could simulate difficult conversations just prior to actual encounters, offering “just-in-time” training \cite{easton2023}. Finally, beyond merely representing the patient, AI may be employed in providing automated feedback to practitioners \cite{Burry2024}.

From a technical perspective, our current platform could benefit from several enhancements. Our open-source lip synchronization is computationally intense, resulting in delays of up to 30 seconds awaiting patient response, which can significantly disrupt the fluidity of conversations. Optimization of software performance and improvement of infrastructure will likely improve this response time. In addition, though our platform can take audio inputs and produce audiovisual outputs, these are translated through an intermediary layer of text (\autoref{technical_flow_diagram}). Recent native multimodal models, such as GPT4-Omni, are trained directly on both text and human voice, and thus are able to better interpret and mimic human vocal cues, such as intonation or pauses. These models may be a superior option for future application development.  

From a research perspective, our approach must be evaluated for educational effectiveness through rigorous methods similar to those established to evaluate the utility of virtual patients. Whether a significant improvement in fidelity of the simulated conversation results in improvement in trainee confidence or performance has yet to be fully evaluated. Qualitative and quantitative explorations of these topics are needed.

Finally, while this work shows promise, we must also carefully consider the ethical implications of using these tools. Beyond addressing copyright issues, reinforcement of biases, or environmental concerns, it is crucial to recognize that standardized patients, primarily actors, depend on collaborations with medical schools for their livelihood. Importantly, though AI-enabled solutions will continue to improve, they cannot replicate the invaluable experience of human interaction. While implementing this these technologies, programs must recognize the substantial contribution of standardized patient actors to medical education and the practice of medicine as a whole. 
 
\section{Acknowledgements, Funding, Data}
The authors wish to thank Larry F. Chu MD, MSc, MS for mentorship and guidance and Lauryn Porte for critical review of the manuscript. S.N.C. was supported by the National Institutes of Health Grant Number T32AI125222.  The content is solely the responsibility of the authors and does not necessarily represent the official views of the National Institutes of Health.

The project's code, patient profiles, and a containerized version of the application have been deposited in a HuggingFace repository and are available at \href{https://doi.org/10.57967/hf/2338}{https://doi.org/10.57967/hf/2338}.


\bibliography{references.bib}

\clearpage\clearpage

\section*{Appendix 1: Technical Implementation Methods} \label{appendix1}
In this study, we present a novel approach to simulating difficult conversations using realistic AI avatars. This intervention involved three primary steps: construction of patient narrative histories, generation of patient multimedia, and integration with a custom video chat application. This document describes the methodologies of these section in greater technical detail. 

\subsection*{Construction of patient profiles}
As described in the primary manuscript, patient profiles were generated based on fictional clinical scenarios and patients. These were combined with a common set of instructions to form the overall prompt, and this prompt was passed to a language model to generate the synthetic patient chatbots. From a technical perspective, we utilized the GPT4 Assistant API (Open AI, San Francisco, CA). Temperature was set at 0.8 to generate a variety of responses. For initial testing, we used the custom GPT feature, which has similar functionality to the Assistant API but allowed for more rapid prototyping. 
 
\subsection*{Image generation}
To generate images and videos of each synthetic patient, multiple imaging models were employed. After developing the patient profiles, we generated images for the patients using Midjourney (version 5, San Francisco, CA) and Stable Diffusion XL (Stability AI, San Francisco, CA). To refine images while maintaining the continuity of the character, we used generative infilling and extracted physical characteristics (via low-rank adaptations, or LoRAs) within Midjourney. Additional fine details were added with Photoshop (Adobe, San Francisco, CA).

\subsection*{Audio generation}
To develop the voices for synthetic patients, we identified royalty-free audio clips of individuals speaking. We collated these files and cleaned them in audio processing software (Ableton Live, Berlin, Germany) and cloned them using ElevenLabs voice cloning tool (ElevenLabs, New York, New York). Manual iterative adjustments were made to the stability, similarity, and style parameters until voices were stable and realistic.

\subsection*{Video generation}
To generate video of the patient responding to queries, we started with the images created above. These were then used as inputs to multiple video-generation tools, including Pika Labs (Pika Labs, Palo Alto, CA), Runway Gen2 (Runway AI, New York, NY), and Stable Diffusion Video (Stability AI, San Francisco, CA) to generate body/head sway movements. These models produced realistic movements of patient’s heads and bodies, but distorted the patients’ face significantly. 

To rectify this, we pivoted to using lip-syncing-focused software for the facial animations. We tested options from D-ID (Tel Aviv, Israel), Synthetsia (London, UK), and HeyGen (Los Angeles, CA), finding that HeyGen produced the most realistic animations, though generation times were substantial (10-20min). This prolonged generation time would limit the ability of these tools to serve as “real time” applications. However, HeyGen does offer a real-time avatar option, though the processing requirements are substantial and thus costs associated with this plan were beyond the budget of this project. Notably, our demonstration video (Supplemental Data Appendix 2) utilizes HeyGen lip-syncronization technology, and has been edited to demonstrate what a real-time avatar may offer. 

Alternative open source lip-syncing models were explored to allow for increased speed. We evaluated wav2lip and its improved versions such as wav2lip with generative adversarial network (GAN), and sync1.6 (SyncLabs, San Francisco, CA), finding that the wav2lip+GAN version had adequate quality with reasonable generation times (20-30 seconds on a consumer-grade desktop). 

A drawback to most of the lip-syncing models was a relative lack of body animation, making the avatars appear to be lifeless “talking heads.” To resolve this, we combined the body sway videos generated with Runway Gen2 with the lipsyncing footage from HeyGen via a feathered overlay in Adobe Premiere (Adobe, San Francisco, CA). This produced a video of the patient subtly shifting their body and speaking some example text with high-quality lip-syncing. This video then served as our base video on which to perform dynamic lip-syncing via the web application.

\subsection*{Web application}
To facilitate real-time interaction with the synthetic patient, we developed a simple web application using Python (Python Foundation, Beaverton, OR). Within this application, users access a mock telehealth interface displaying a video feed of the patient awaiting the provider’s question. Users initiate communication by pressing a speech button, allowing their question to be recorded. This audio is subsequently transcribed into text using a voice-to-text model, WhisperAPI (Open AI, San Francisco, CA). The text is then sent to the OpenAI inference API, which generates a text response. The response is then converted into audio using the patient’s predetermined voice. To lip-sync the audio to the patient video, the aformentioned base video is used as the template. This base video and the generated audio are passed to the lipsync engine. This generates an audiovisual clip that appears as if the patient is directly responding to the provider's inquiry. This clip is then integrated into the live video feed, allowing for a seamless interaction.

To permit individuals to experiment with our platform, we developed a containerized version of the application using Docker. Instructions for installation have been deposited in a HuggingFace repository and are available at \href{https://doi.org/10.57967/hf/2338}{https://doi.org/10.57967/hf/2338}. Though the application size is approximately 5GB, containerization allows the software to be run on almost any desktop. Users will need accounts and API keys from OpenAI and ElevenLabs. Because of the limitations of audio input-output within the containerized application, the audio-to-text functionality is not available. Instead, users will need to provide their questions in the form of text.

\clearpage

\section*{Appendix 2: Example of Encounter with Synthetic Patient} \label{appendix2}

Video file available at \href{https://doi.org/10.6084/m9.figshare.25930861}{https://doi.org/10.6084/m9.figshare.25930861}

\end{document}